\documentclass{elsart}
\usepackage{epsfig}

\begin{document}

\begin{frontmatter}

\title{New limits on nucleon decays into invisible channels
with the BOREXINO Counting Test Facility}

\begin{center}
\author[a]{H.O.~Back},
\author[b]{M.~Balata},
\author[c]{A.~de~Bari},
\author[d]{T.~Beau},
\author[d]{A.~de~Bellefon},
\author[e]{G.~Bellini $^{\clubsuit,}$},
\author[f]{J.~Benziger},
\author[e]{S.~Bonetti},
\author[g]{C.~Buck},
\author[e]{B.~Caccianiga},
% ...

\address[a]{Virginia Polytechnique Institute and State Univercity Blacksburg, VA 24061-0435, Virginia, USA}
\address[b]{L.N.G.S. SS 17 bis Km 18+910, I-67010 Assergi(AQ), Italy}
\address[c]{Dipartimento di Fisica Nucleare e Teorica Universita` di Pavia, Via A. Bassi, 6 I-27100, Pavia, Italy}
\address[d]{Laboratoire de Physique Corpusculaire et Cosmologie, 11 place Marcelin Berthelot 75231 Paris Cedex 05, France}
\address[e]{Dipartimento di Fisica Universit\`a di Milano, Via Celoria, 16 I-20133 Milano, Italy}
\address[f]{Dept. of Physics, Princeton University, Jadwin Hall, Washington Rd, Princeton NJ 08544-0708, USA}
\address[g]{Max-Planck-Institut fuer Kernphysik, Postfach 103 980 D-69029, Heidelberg, Germany}
\address[h]{Dept. of Physics, Queen's University Stirling Hall, Kingston, Ontario K7L 3N6, Canada}
\address[i]{Technische Universitaet Muenchen, James Franck Strasse, E15 D-85747, Garching, Germany}
\address[j]{Joint Institute for Nuclear Research, 141980 Dubna, Russia}
\address[k]{Dept. of Physics Massachusetts Institute of Technology, Cambridge, MA 02139, USA}
\address[l]{Dipartimento di Chimica Universit\`a di Perugia, Via Elce di Sotto, 8 I-06123, Perugia, Italy}
\address[m]{RRC Kurchatov Institute, Kurchatov Sq.1, 123182 Moscow, Russia}
\address[n]{Dipartimento di Fisica Universit\`a and I.N.F.N. Genova, Via Dodecaneso, 33 I-16146 Genova, Italy}
\address[o]{KFKI-RMKI, Konkoly Thege ut 29-33 H-1121 Budapest, Hungary}
\address[p]{Bell Laboratories, Lucent Technologies, Murray Hill, NJ 07974-2070, USA}
\address[q]{M. Smoluchowski Institute of Physics, Jagellonian University, PL-30059 Krakow, Poland}
\address[r]{Institute for Nuclear Research, MSP 03680, Kiev, Ukraine}

\corauth[cor1]{Corresponding author. St. Petersburg Nucl. Phys.
Inst., 188350 Gatchina, Russia. E-mail: derbin@mail.pnpi.spb.ru}
\corauth[cor2]{~Corresponding author. Joint Inst. for Nucl.
Research, 141980 Dubna, Russia. E-mail: smirnov@lngs.infn.it}
\corauth[cor3]{~~Corresponding author. Institute for Nuclear
Research, MSP 03680, Kiev, Ukraine. E-mail: tretyak@lngs.infn.it}

\thanks[lab1]{On leave of absence from St. Petersburg Nuclear Physics Inst. - Gatchina, Russia}
\thanks[lab2]{Marie Curie fellowship at LNGS}
\thanks[lab3]{On leave of absence from Institute for Nuclear Research, MSP 03680, Kiev, Ukraine}
\thanks[lab4]{On leave of absence from KFKI-RMKI, Konkoly Thege ut 29-33 H-1121 Budapest, Hungary}
\thanks[lab6]{Deceased\\ $^{\clubsuit}$Spokesman\\$^{\dagger}$Project manager\\$^{\Box}$Operational manager\\$^{\triangle}$GLIMOS\\$^{\bullet}$Task manager}

\end{center}

\begin{abstract}
The results of background measurements with the second version of
the BOREXINO Counting Test Facility (CTF-II), installed in the
Gran Sasso Underground Laboratory, were used to obtain limits on
the instability of nucleons, bounded in nuclei, for decays into
invisible channels ($inv$): disappearance, decays to neutrinos,
etc. The approach consisted of a search for decays of unstable
nuclides resulting from $N$ and $NN$ decays of parent $^{12}$C,
$^{13}$C and $^{16}$O nuclei in the liquid scintillator and the
water shield of the CTF. Due to the extremely low background and
the large mass (4.2 ton) of the CTF detector, the most stringent
(or competitive) up-to-date experimental bounds have been
established: $\tau(n  \rightarrow inv) > 1.8 \cdot 10^{25}$ y,
$\tau(p \rightarrow inv) > 1.1 \cdot 10^{26}$ y, $\tau(nn
\rightarrow inv)
> 4.9 \cdot 10^{25}$ y and $\tau(pp \rightarrow inv) > 5.0 \cdot
10^{25}$ y, all at 90\% C.L.
\end{abstract}


\end{frontmatter}

\section{Introduction}

The baryon ($B$) and lepton ($L$) numbers are considered to be
conserved in the Standard Model (SM)\footnote{It should be noted
that nonperturbative effects at high energies can lead to the $B$
and $L$ violation even in the SM \cite{Hoo76}.}. However, no
symmetry principle underlies these laws, such as, e.g. gauge
invariance, which guarantees conservation of the electric charge.
Many extensions of the SM include $B$ and $L$ violating
interactions, predicting the decay of protons and neutrons bounded
in nuclei. Various decay mechanisms with $\Delta B$=1, 2 and
$\Delta (B$--$L)$=0, 2 have been discussed in the literature
intensively \cite{Lan81,Car01}. A novel baryon number violating
process, in which two neutrons in a nucleus disappear, emitting a
bulk majoron $nn \rightarrow \chi$, was proposed recently
\cite{Moh00}; the expected mean lifetime was estimated to be
$\sim$10$^{32-39}$ y. Additional possibilities for the nucleon
($N$) decays are related to theories which describe our world as a
brane inside higher-dimensional space \cite{Ynd91,Dub00}.
Particles, initially confined to the brane, may escape to extra
dimensions, thus disappearing for the normal observer; the
characteristic proton mean lifetime was calculated to be
$\tau(p)$=9.2$\cdot10^{34}$ y \cite{Dub02}. Observation of the
disappearance of $e^{-}$, $N$, $NN$ would be a manifestation of
the existence of such extra dimensions \cite{Dub00}.

No evidence for nucleon instability has been found to date.
Experimental searches \cite{Per84} with the IMB, Fr\'ejus,
(Super)Kamiokande and other detectors have been devoted mainly to
nucleon decays into strongly or electromagnetically interacting
particles, where lower limits on the nucleon mean lifetime of
$10^{30-33}$ y were obtained \cite{PDG00}. At the same time, for
modes where $N$ or $NN$ pairs disappear or they decay to some
weakly interacting particles (neutrinos, majorons, etc.), the
experimental bounds are a few orders of magnitude lower. Different
methods were applied to set limits for such decays\footnote{We are
using the following classification of decay channels: decay to
invisible channel ($inv$) means disappearance or decay to weakly
interacting particles (one or few neutrinos of any flavors,
majorons, etc.). Channel to $anything$, mentioned below, includes
also decays to $inv$.} (see table 1 for summary):

\def\arraystretch{1.05}

\begin{table}[!htbp]
\begin{center}
\caption{Lower limits on the mean lifetime for decay of nucleons,
bounded in nuclei, into invisible channels established in various
approaches (see footnote 7 for channels classification).}
\begin{tabular}{|llll|}
\hline
\multicolumn{2}{|l}{$N$ or $NN$ decay}                    & $\tau$ limit, y         & Year, reference and short explanation \\
~    & ~                                                  & and C.L.                & \\
\hline
$p$  & $\rightarrow$ $anything$                           & 1.2$\cdot$10$^{23}$     & 1958 \cite{Fle58} limit on $^{232}$Th spontaneous fission \\
~    & ~ & 3.0$\cdot$10$^{23}$      & 1970
\cite{Dix70} search for free $n$ in liquid scintillator \\ ~    &
~ & ~                        & ~~~~~~~~~~~~ enriched in deuterium
($d \rightarrow n+?$) \\ ~    & ~ & 4.0$\cdot$10$^{23}$ 95\% &
2001 \cite{Tre01} free $n$ in D$_2$O volume \\ ~    &
$\rightarrow$ $inv$ & 7.4$\cdot$10$^{24}$      & 1977 \cite{Eva77}
geochem. search for
$^{130}$Te$\rightarrow$...$\rightarrow$$^{129}$Xe \\ ~    & ~ &
1.1$\cdot$10$^{26}$      & 1978 \cite{Fir78} radiochem. search for
$^{39}$K$\rightarrow$...$\rightarrow$$^{37}$Ar \\ ~    & ~ &
1.9$\cdot$10$^{24}$ 90\% & 2000 \cite{Ber00} search for $^{128}$I
decay in $^{129}$Xe detector \\ ~    & ~ & $\sim$10$^{28}$ ~ &
2002 \cite{SNO-NC} free $n$ in D$_2$O volume \\ ~    & ~ &
3.5$\cdot$10$^{28}$ 90\% & 2003 \cite{Zde02} free $n$ in D$_2$O
volume
\\ ~    & ~ & 1.1$\cdot$10$^{26}$ 90\% & 2003 [~$^a$~] search for
$^{12}$B decay in the CTF detector \\
$n$  & $\rightarrow$ $anything$                           &
1.8$\cdot$10$^{23}$      & 1958 \cite{Fle58} limit on $^{232}$Th
spontaneous fission \\ ~    & $\rightarrow$ $inv$ &
8.6$\cdot$10$^{24}$      & 1977 \cite{Eva77} geochem. search for
$^{130}$Te$\rightarrow$...$\rightarrow$$^{129}$Xe \\ ~    & ~ &
1.1$\cdot$10$^{26}$      & 1978 \cite{Fir78} radiochem. search for
$^{39}$K$\rightarrow$...$\rightarrow$$^{37}$Ar \\ ~    & ~ &
4.9$\cdot$10$^{26}$ 90\% & 1993 \cite{Suz93} search for $\gamma$
with $E_\gamma$=19--50 MeV emitted in \\ ~    & ~ & ~
& ~~~~~~~~~~~~ $^{15}$O deexcitation in Kamiokande detector \\ ~
& ~ & 1.8$\cdot$10$^{25}$ 90\% & 2003 [~$^a$~] search for $^{11}$C
decay in the CTF \\ ~    & $\rightarrow$ $\nu_\mu
\overline{\nu}_\mu \nu_\mu$ & 5.0$\cdot$10$^{26}$ 90\% & 1979
\cite{Lea79} massive liquid scint. detector fired by $\nu_\mu$ \\
~    & ~                                                  & ~ &
~~~~~~~~~~~~ in result of $n$ decays in the whole Earth $^{b,c}$
\\ ~    & ~                                                  &
1.2$\cdot$10$^{26}$ 90\% & 1991 \cite{Ber91} Fr\'{e}jus iron
detector fired by $\nu_\mu$ $^c$ \\ ~    & $\rightarrow$ $\nu_e
\overline{\nu}_e \nu_e$       & 3.0$\cdot$10$^{25}$ 90\% & 1991
\cite{Ber91} Fr\'{e}jus iron detector fired by $\nu_e$ $^d$ \\ ~
& $\rightarrow$ $\nu_i \overline{\nu}_i \nu_i$       &
2.3$\cdot$10$^{27}$ 90\% & 1997 \cite{Gli97} search for bremsstr.
$\gamma$ with $E_\gamma$$>$100 MeV \\ ~    & ~
& ~                        & ~~~~~~~~~~~~ emitted due to sudden
disapp. of $n$ magn. \\ ~    & ~
& ~                        & ~~~~~~~~~~~~ moment (from Kamiokande
data) $^e$ \\ ~    & $\rightarrow$
$\nu_i\overline{\nu}_i\nu_i\overline{\nu}_i\nu_i$ &
1.7$\cdot$10$^{27}$ 90\% & 1997 \cite{Gli97} the same approach
$^e$ \\
$nn$ & $\rightarrow$ $\nu_\mu \overline{\nu}_\mu$         &
6.0$\cdot$10$^{24}$ 90\% & 1991 \cite{Ber91} Fr\'{e}jus iron
detector fired by $\nu_\mu$ $^f$ \\ ~    & $\rightarrow$ $\nu_e
\overline{\nu}_e$             & 1.2$\cdot$10$^{25}$ 90\% & 1991
\cite{Ber91} Fr\'{e}jus iron detector fired by $\nu_e$ $^g$ \\ ~ &
$\rightarrow$ $inv$                                &
1.2$\cdot$10$^{25}$ 90\% & 2000 \cite{Ber00} search for $^{127}$Xe
decay in $^{129}$Xe detector \\ ~    & ~ & 4.2$\cdot$10$^{25}$
90\% & 2002 \cite{Bel03} radiochem. search for
$^{39}$K$\rightarrow$...$\rightarrow$$^{37}$Ar $^h$ \\ ~    & ~
& 4.9$\cdot$10$^{25}$ 90\% & 2003 [~$^a$~] search for $^{10}$C and
$^{14}$O decay in the CTF \\
$pp$ & $\rightarrow$ $inv$                                &
5.5$\cdot$10$^{23}$ 90\% & 2000 \cite{Ber00} search for $^{127}$Te
decay in $^{129}$Xe detector \\ ~    & ~ & 5.0$\cdot$10$^{25}$
90\% & 2003 [~$^a$~] search for $^{11}$Be decay in the CTF
detector \\ $pn$ & $\rightarrow$ $inv$ & 2.1$\cdot$10$^{25}$ 90\%
& 2002 \cite{Bel03} radiochem. search for
$^{39}$K$\rightarrow$...$\rightarrow$$^{37}$Ar $^h$ \\
\hline
\multicolumn{4}{l}{$^a$ This work} \\ \multicolumn{4}{l}{$^b$ The
result of \cite{Lea79} was reestimated in \cite{Ber91} to be more
than one order of magnitude lower} \\ \multicolumn{4}{l}{$^c$ The
limit is also valid for $p$ $\rightarrow$ $\nu_\mu
\overline{\nu}_\mu \nu_\mu$ decay} \\ \multicolumn{4}{l}{$^d$ The
limit is also valid for $p$ $\rightarrow$ $\nu_e \overline{\nu}_e
\nu_e$ decay} \\ \multicolumn{4}{l}{$^e$ $i=e, \mu, \tau$} \\
\multicolumn{4}{l}{$^f$ The limit is also valid for $pn$ and $pp$
decays into $\nu_\mu \overline{\nu}_\mu$} \\
\multicolumn{4}{l}{$^g$ The limit is also valid for $pn$ and $pp$
decays into $\nu_e \overline{\nu}_e$} \\ \multicolumn{4}{l}{$^h$
On the base of the data of ref. \cite{Fir78}} \\
\end{tabular}
\end{center}
\end{table}

(1) Using the limit on the branching ratio of spontaneous fission
of $^{232}$Th under the assumption that $p$ or $n$ decay in
$^{232}$Th will destroy the nucleus \cite{Fle58}. The bound on the
mean lifetime obtained in this way can be considered independent
of the a $p$ or $n$ decay mode, since the $^{232}$Th nucleus can
be destroyed either by strong or electromagnetic interactions of
daughter particles with the nucleus or, in the case of $N$
disappearance, by subsequent nuclear deexcitation process;

(2) Search for a free $n$ created after $p$ decay or disappearance
in the deuterium nucleus ($d$=$pn$) in a liquid scintillator
enriched in deuterium \cite{Dix70} or in a volume of D$_2$O
\cite{Tre01,SNO-NC,Zde02};

(3) Geochemical \cite{Eva77} or radiochemical \cite{Fir78} search
for daughter nuclides which have appeared after $N$ decays in the
mother nuclei (valid for decays into invisible channels);

(4) Search for prompt $\gamma $ quanta emitted by a nucleus in a
de-excitation process after $N$ decays within the inner nuclear
shell \cite{Suz93} (valid for invisible channels);

(5) Considering the Earth as a target with nucleons which decay by emitting
electron or muon neutrinos; the $\nu_e$, $\nu_\mu$ can be detected by a large
underground detector \cite{Lea79,Ber91} (valid for decay
into neutrinos with specific flavors);

(6) Search for bremsstrahlung $\gamma$ quanta emitted due to a sudden
disappearance of the neutron magnetic moment \cite{Gli97} (limits depend on
the number of emitted neutrinos);

(7) Study of radioactive decay of daughters (time-resolved from prompt
products), created as a result of $N$ or $NN$ decays of the mother nuclei,
incorporated into a low-background detector (valid for decay
into invisible channels). This method was first exploited by the
DAMA group with a liquid Xe detector \cite{Ber00}.

In the present paper we use the same approach to search for $N$
and $NN$ instability with the Counting Test Facility, a 4.2 ton
prototype of the multiton BOREXINO detector for low energy solar
neutrino spectroscopy \cite{BORgen}. The preliminary results were
presented in \cite{N2002}.

\section{Experimental set-up and measurements}


\subsection {Technical information about CTF and BOREXINO}

BOREXINO, a real-time 300 ton detector for low-energy neutrino spectroscopy,
is nearing completion in the Gran Sasso Underground Laboratory
(see \cite{BORgen} and refs. therein). The main goal of the detector is the
measurement of the $^7$Be solar neutrino flux via $\nu-e$ scattering in an ultra-pure
liquid scintillator, while several other basic questions in astro- and particle
physics will also be addressed.

The Counting Test Facility (CTF), installed in the Gran Sasso Underground
Laboratory, is a prototype of the BOREXINO detector.
Detailed reports on the CTF results
have been published \cite{BORgen,CTFgen,CTFlgt},
and only the main characteristics of the set-up are outlined here.

The CTF consists of an external cylindrical water tank ($\oslash
$11$\times $10 m; $\approx$1000 t of water) serving as passive
shielding for 4.2 m$^3$ of liquid scintillator (LS) contained in
an inner spherical vessel of $\oslash $2.0 m. High purity water
with a radio-purity of $\approx $$10^{-14}$ g/g (U, Th), $\approx
$$10^{-12}$ g/g (K) and $<$2 $\mu $Bq/{\it l} for $^{222}$Rn is
used for the shielding. The LS was purified to the level of
$\simeq$10$^{-16}$ g/g in U/Th contamination.

We analyze here the data following the upgrade of the CTF (CTF-II). The
liquid scintillator used at this stage was a phenylxylylethane
(PXE, C$_{16}$H$_{18}$) with p-diphenylbenzene (para-terphenyl) as
a primary wavelength shifter at a concentration of 2 g/{\it l}
along with a secondary wavelength shifter
1,4-bis-(2-methylstyrol)-benzene (bis-MSB) at 50 mg/{\it l}. The
density of the scintillator is 0.996 kg/{\it l}. The scintillator
principal de-excitation time is less than 5 ns which provides a good
position reconstruction. In the CTF-II an additional nylon screen
between the scintillator vessel and PMTs (against radon
penetration) and a muon veto system were installed.

The scintillation light is collected with 100 phototubes (PMT) fixed
to a 7 m diameter support structure inside the water tank. The PMTs are
fitted with light concentrators which
provide a total of 21\% optical coverage. The number of photoelectrons measured
experimentally is 3.54 per PMT for 1 MeV electrons
at the detector's center.

For each event the charge and timing of hit PMTs are
recorded. Each channel is supported by an auxiliary channel, the
so-called second group of electronics, used to record all
events coming within a time window of 8.2 ms after the trigger, which allows
tagging of fast time-correlated events with a decrease of the overall
dead time of the detector. For longer delays, the computer
clock is used providing the accuracy of $\approx$0.1 s. Event
parameters measured in the CTF include the total charge
collected by the PMTs during 0--500 ns, used to determine an event's energy;
the charge in the "tail" of the pulse (48--548 ns) which is used for pulse
shape discrimination;
 PMT timing, used to reconstruct the event's position; and the time
elapsed between sequential events, used to tag time-correlated events.

\subsection{Detector calibration}
  The energy of an event in the CTF detector is defined using the total
collected charge from all PMT's. In a simple approach the energy
is supposed to be linear with respect to the total collected charge.
The coefficient linking the event energy and the total collected charge
is called light yield (or photoelectron yield). The light yield for
electrons can be considered linear with respect to its energy only
for energies above 1 MeV. At low energies the phenomenon
of ``ionization quenching'' violates the linear dependence
of the light yield versus energy \cite{Birks}. The deviations from the
linear law can be taken into account by the ionization deficit function
\( f(k_{B},E) \), where $k_B$ is an empirical Birks' constant.
For the calculations of the ionitazion quenching effect for PXE scintillator
we used the KB program from the CPC library \cite{Quenching}.
  The ``ionization quenching'' effect leads to a shift in the position
of the full energy peak for gammas on the energy scale calibrated
using electrons. In fact, the position of the 1461 keV $^{40}$K
gamma in the CTF-II data corresponds to 1360 keV of energy
deposited for an electron.

  The detector energy and spatial resolution were studied with
radioactive sources placed at different positions inside the
active volume of the CTF. A typical spatial 1$\sigma$ resolution
is 10~cm at 1 MeV. The studies showed also that the total charge
response of the CTF detector can be approximated by
 a Gaussian.  For energies $E$$\ge$1 MeV
(which are of interest here),
 the relative resolution can be expressed as
$\sigma_E/E=\sqrt{3.8\cdot10^{-3}/E+2.3\cdot10^{-3}}$ ($E$ is in MeV)
\cite{Smi00} for events uniformly distributed over the detector's volume.

  The energy dependence on the collected charge becomes non-linear for
the energies $E\simeq4.5$~MeV in the first group of electronics because
of the saturation of the ADCs used. In this region we are using only the fact of
observing or not-observing candidate events, hence the mentioned
nonlinearity doesn't influence the result of the analysis.

Further details on the energy and spatial resolutions of the detector
and ionization quenching for electrons, $\gamma$ quanta and $\alpha$
particles can be found in \cite{Smi00,BORel}.

\subsection{Muon veto}

The upgrade of the CTF was equipped with a carefully designed muon veto system.
It consists of 2 rings of 8 PMTs each, installed at the bottom of the tank.
The radii of the rings are 2.4 and 4.8 m. Muon veto PMTs are
looking upward and have no light concentrators.
The muon veto system was optimized in order to have a negligible probability
of registering the scintillation events in the so-called "neutrino energy
window" (250--800 keV). The behaviour of the muon veto at the higher
energies has been specially studied for the present work.
Experimental measurements with a radioactive source (chain of $^{226}$Ra)
\cite{CTFRn} gave, for the probability $\eta(E)$ of identification of an event
with energy $E$ in the LS by the muon veto, the value of (1$\pm$0.2)\%
in the 1.8--2.0 MeV region (see fig. 3 later).
The energy dependence of $\eta(E)$ was also
calculated by a ray-tracing Monte Carlo method accounting
for specific features of the light propagation in the CTF which
are detailed in \cite{CTFlgt}. The calculated function was
adjusted to reproduce correctly the experimental measurements with the
$^{226}$Ra source.

\subsection{Data selection}

As it will be shown below, the candidate events, relevant for our
studies, have to satisfy the following criteria:
(1) the event should occur in the active volume of the detector and must not
be accompanied by the muon veto tag;
(2) it should be single (not followed by a time-correlated event);
(3) its pulse shape must correspond to that of events caused by
$\gamma$ or $\beta$ particles.

The selection and treatment of data (spatial cuts, analysis of an event's pulse shape
to distinguish between electrons and $\alpha$ particles, suppression of
external background by the muon veto system, etc.) is similar to that
in ref. \cite{BORel}.

The experimental energy spectra in CTF-II,
accumulated during 29.1 days of measurements, are shown in fig. 1. The
spectrum without any cuts (spectrum 1) is presented on the top.
The second spectrum is obtained by applying the muon cut, which
suppressed the background rate by up to two orders of magnitude, depending
on the energy region.

\nopagebreak
\begin{figure}[htbp]
\begin{center}
\mbox{\epsfig{figure=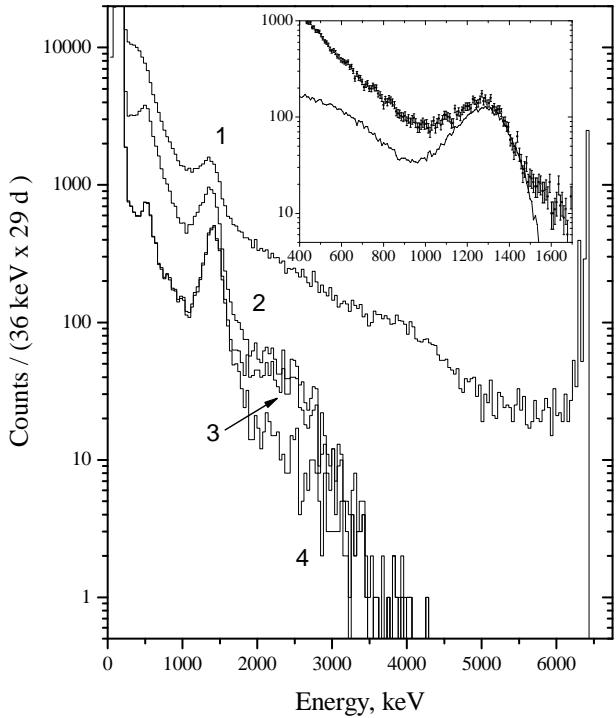,width=9.0cm}}
\caption{Background energy spectra of the 4.2 ton BOREXINO
CTF-II detector measured during 29.1 days. From top to bottom: (1)
spectrum without any cuts; (2) with muon veto; (3) only events
inside the radius $R$$\leq$100 cm with additional $\alpha/\beta$
discrimination applied to eliminate any contribution from $\alpha$
particles; (4) pairs of correlated events (with time interval
$\Delta t$$\leq$8.2 ms between signals) are removed. In the inset,
the simulated response function for external $^{40}$K gammas is
shown together with the experimental data.}
\end{center}
\end{figure}

On the next stage of the data selection we applied a cut on the reconstructed radius.
In the energy region 1--2 MeV we used $R\leq$100 cm cut aiming to
remove the surface background events (mainly due to the $^{40}$K
decays outside the inner vessel) and leave the events uniformly
distributed over the detector volume. The efficiency of the cut
has been studied with MC simulation and lays in the range of
$\epsilon_R=0.76-0.80$ in the energy region 1--2 MeV. Additional
$\alpha /\beta$ discrimination \cite{CTFgen} was applied to
eliminate contribution from $\alpha$ particles (spectrum 3 in fig.
1).

The time-correlated events (that occurred in the time window
$\Delta t$$<$8.2 ms) were also removed (spectrum 4).
 The peak at 1.46 MeV, present in all spectra, is due to $^{40}$K
decays outside the scintillator, mainly in the ropes supporting
the nylon sphere.
 The peak-like structure at $\sim$6.2 MeV
is caused by saturation of the electronics
by high-energy events.
The lower spectrum of fig. 1 presents all
the candidate scintillation events in the search for the decay of radioactive
nuclides created in the active volume of the CTF after the nucleon
disappearance in parent nuclei.

\section{Data analysis and results}

\subsection{Theoretical considerations}

The decay characteristics of the daughter nuclides, resulting from
$N$ and $NN$ decays in parent nuclei --
$^{12}$C, $^{13}$C and $^{16}$O -- contained in the sensitive
volume of the CTF liquid scintillator or in the water shield,
are listed in table 2.

\begin{table}[!htbp]
\begin{center}
\caption{Initial nuclei in the Counting Test Facility, their
abundance $\delta$ \cite{Ros98}, processes with $\Delta B=1,2$ and
characteristics of daughter nuclides \cite{TOI78}.~}
\begin{tabular}{|lllll|}
\hline Initial                   & De-  &
\multicolumn{3}{l|}{Daughter nucleus, half-life, modes of decay
and energy release}          \\ nucleus                   & cay  &
&                              &
\\ \hline ~ & & & & \\ $^{12}_{~6}$C             & $n$  &
$^{11}_{~6}$C  & $T_{1/2}$=20.38 m             &  $\beta^+$
(99.76\%), EC(0.24\%); $Q$=1.982 MeV \\ $\delta$=$98.93$\%
& $p$  & $^{11}_{~5}$B  & stable                        &
\\ ~                         & $nn$ & $^{10}_{~6}$C  &
$T_{1/2}$=19.2 s              & $\beta^+$; $Q$=3.651 MeV
\\ ~                         & $pn$ & $^{10}_{~5}$B  & stable
&                                                 \\ ~
& $pp$ & $^{10}_{~4}$Be & $T_{1/2}$=1.6$\cdot$10$^6$ y &
$\beta^-$; $Q$=0.556 MeV                        \\ ~ & & & & \\
$^{13}_{~6}$C             & $n$  & $^{12}_{~6}$C  & stable
&                                                 \\
$\delta$=$1.07$\%         & $p$  & $^{12}_{~5}$B  & $T_{1/2}$=20.4
ms             & $\beta^-$; $Q$=13.370 MeV
\\ ~                         & $nn$ & $^{11}_{~6}$C  &
$T_{1/2}$=20.38 m             &  $\beta^+$ (99.76\%), EC(0.24\%);
$Q$=1.982 MeV \\ ~                         & $pn$ & $^{11}_{~5}$B
& stable                        &
\\ ~                         & $pp$ & $^{11}_{~4}$Be &
$T_{1/2}$=13.8 s              & $\beta^-$; $Q$=11.508 MeV
\\ ~ & & & & \\ $^{16}_{~8}$O             & $n$  & $^{15}_{~8}$O
& $T_{1/2}$=122 s               &  $\beta^+$ (99.89\%),
EC(0.11\%); $Q$=2.754 MeV \\ $\delta$=$99.757$\%       & $p$  &
$^{15}_{~7}$N  & stable                        &
\\ ~                         & $nn$ & $^{14}_{~8}$O  &
$T_{1/2}$=70.60 s             & $\beta^+$; $Q$=5.145 MeV
\\ ~                         & $pn$ & $^{14}_{~7}$N  & stable
&                                                 \\ ~
& $pp$ & $^{14}_{~6}$C & $T_{1/2}$=5730 y              &
$\beta^-$; $Q$=0.156 MeV                        \\ ~ & & & & \\
\hline
\end{tabular}
\end{center}
\end{table}

After the disappearance of one or two nucleons in the parent
nuclide, one or two holes appear in the nuclear shells; these
holes will be filled in a subsequent nuclear de-excitation
process, unless the nucleons reside on the outermost shells. If
the initial excitation energy, $E_{exc}$, is higher than the
binding energy $S_N$ of the least bound nucleon ($N$ is $p$ or
$n$), the nucleus will be de-excited by particle emission ($p$,
$n$, $\alpha$, $d$, etc.); otherwise ($E_{exc}<S_N$) a $\gamma$
quantum will be emitted. In the following, we will take into
consideration the $N$ and $NN$ decays from the last filled
single-particle levels, when only $\gamma$ quanta could be
emitted. Thus the daughter nucleus is exactly known. We also
neglect by the prompt signal from gamma quanta emitted in initial
deexcitation process (in particular, because of generally big half
lives of daughter nuclei, and uncertainty in the number of emitted
$\gamma$'s, their energies, and rejection of high energy events by
the muon veto), and search only for subsequent radioactive decay
of created nuclei.

The number of nucleons or nucleon pairs participating in the process
can be calculated following refs. \cite{Eva77,Ber00}. For example,
after the decay of a
neutron with binding energy $E^b_n(A,Z)$, the excitation energy of the
$(A-1,Z)$ daughter nucleus will be $E_{exc}=E^b_n(A,Z)-S_n(A,Z)$. The
condition to emit only $\gamma$ quanta in the deexcitation process,
$E_{exc}<S_N(A-1,Z)$, gives this restriction on the neutron binding energy:
$E^b_n(A,Z)<S_n(A,Z)+S_N(A-1,Z).$ Similar equations can be written for $p$,
$pp$, $pn$ and $nn$ decays (see ref. \cite{Ber00}).
The values of the separation energies $S_N$ and $S_{NN}$ were taken from
ref. \cite{Aud95}. For single-particle energies of nucleons $E_{p,n}^b$ on
nuclear shells we used the continuum shell model calculations \cite{Fri75}
for $^{12}$C and $^{13}$C, and the Hartree-Fock calculations with the Skyrme's
interaction \cite{Vau72} for $^{16}$O.

\subsection{Simulation of the response functions}

The expected response functions of the CTF detector and related efficiencies
for the decay of unstable
 daughter nuclei were simulated with the EGS4
package \cite{EGS4}.
The number of initial electrons and $\gamma$ quanta
emitted in the decay of the nucleus and their energies
were generated according to the decay schemes \cite{TOI78}.
The events were supposed to be uniformly distributed in the whole volume of the
liquid scintillator (and in a water layer close to the LS).
The energy and spatial resolution of the detector \cite{Smi00},
light quenching factors for electrons and gammas \cite{Quenching},
$\mu$ veto and triggering efficiency were taken into
account in the simulations.
The calculated responses for the decay of $^{11}$C and $^{10}$C in the
liquid scintillator (created after $n$ and $nn$
disappearance in $^{12}$C, respectively) and $^{14}$O in the water
shield ($nn$ decay in $^{16}$O) are shown in fig. 2. In the last
case only the water layer of 1 m thickness
closest to the liquid scintillator  was taken into consideration.
Contributions from layers further out are negligible.

\nopagebreak
\begin{figure}[htbp]
\begin{center}
\mbox{\epsfig{figure=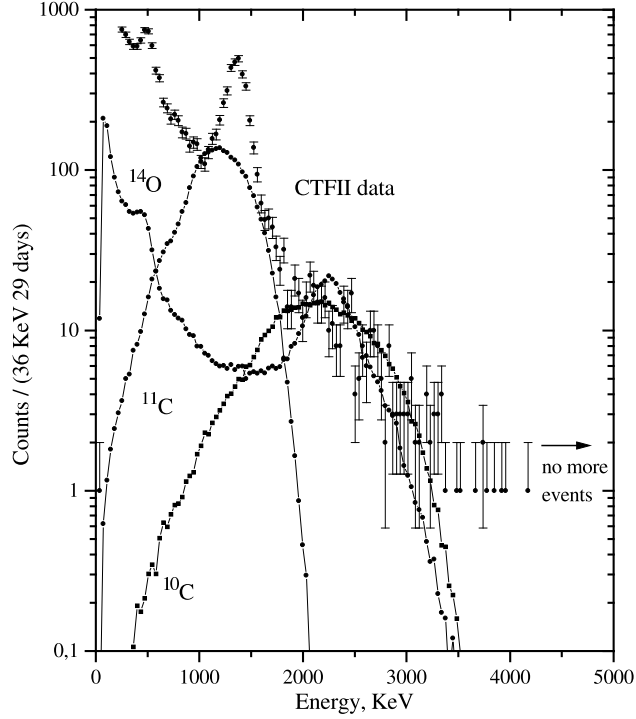,width=9.0cm}}
\caption{Energy distribution of the CTF-II installation
collected during 29.1 d with all cuts. The expected response
functions of the detector are also shown for $^{11}$C
($3.4\cdot10^3$ decays in the liquid
 scintillator;
corresponding mean lifetime for the $n$ decay is
$\tau_{n}=1.8\cdot 10^{25}$ y), $^{10}$C ($6.8\cdot10^2$ decays;
$\tau_{nn}=4.4\cdot 10^{25}$ y), and $^{14}$O ($1.4\cdot10^4$ decays in 1 m thick
water layer closest to the sphere with liquid scintillator;
$\tau_{nn}=5.7\cdot 10^{24}$ y).}
\end{center}
\end{figure}

  The CTF background was simulated as well in order to check the
understanding of the detector. The simulated $^{40}$K 1.46~MeV
gamma peak together with experimental data is shown in the inset
of fig. 1. The shift in the position of the full absorption peak
is due to the ionization quenching effect (see section 2.2). One
can see the good agreement between the experimental and simulated
$^{40}$K peak position and resolution.


\subsection{Limits on probabilities of the $N$ and $NN$ disappearance}

The experimental data (see fig. 2, where the spectrum with all cuts
is shown in detail) gives no strong evidence of
the expected $N$ and $NN$ decay response functions, thus allowing
only bounds to be set on the processes being searched for.
In the present study, in order
to extract the limits on the relevant mean lifetimes,
we assumed conservatively that {\it all} events in the
CTF experimental spectrum in the corresponding energy range
$\Delta E$ are due to nucleon decays.

The mean lifetime limit was estimated using the formula

\begin{equation}
\tau _{\lim}=\varepsilon _{\Delta E}\cdot N_{nucl}\cdot N_{obj}\cdot
t/S_{\lim } = N_{nucl}\cdot N_{obj}\cdot t/D_{\lim },
\end{equation}

\noindent where $\varepsilon _{\Delta E}$ is the detection
efficiency in the $\Delta E$ energy window calculated in the full
simulation of the relevant process, taking into account the radial cut
efficiency $\epsilon_R$, and probability of identification by muon
veto $\eta$; $N_{nucl}$ the number of parent nuclei; $N_{obj}$ the
number of objects ($n$, $p$ or $NN$ pairs) inside the parent
nucleus whose decay will give the specific daughter nucleus; $t$
the time of measurements; $S_{\lim}$ the number of events (in the
$\Delta E$ window) due to a particular effect which can be
excluded with a given confidence level on the basis of
experimental data; and $D_{\lim}=S_{\lim}/\varepsilon _{\Delta E}$
the corresponding number of decays in the liquid
scintillator/water.

 The parameters in (1) were defined in the following way. The mass of
scintillator was defined measuring the buoyancy with a precision
of 5$\%$. The total time of the data taking is 29.1 days and takes
into account the dead time of the electronics of $\simeq3\%$. The
measurements were performed in runs of about 24 hours. The
internal source of $^{40}$K was used to check the energy calibration
stability. The position of the $^{40}$K peak was defined for every
run with statistical accuracy of $\simeq20$ keV. No systematical
shift in energy scale with a time has been found. The same is
valid for the energy resolution, which is directly connected to
the energy scale. In addition, all the response functions for the
searched decays are wider than detector's resolution, hence the
result is insensitive to the observed energy scale uncertainties.

  The probability of detecting scintillation events by the muon veto
$\eta=1\pm0.2\%$ has been defined from the experimental data for
the energy $E=1.9$~MeV. This value is in agreement with the MC
simulation. For $\eta(1.9$ MeV$)=1.2\%$ the changes in the
integral efficiency $1-\eta(E)$ is negligible in the case of $n$
and $nn$ decays, for the $p$ and $pp$ decays the
$\varepsilon_{\Delta E}$ will decrease by 7$\%$.

\subsubsection{$n$ and $nn$ disappearance}

For the $nn$ decay, there are four neutrons on the outermost
$1p_{3/2}$ level of $^{12}$C which gives the number of $nn$ pairs
$N_{obj}=2$. The disappearance of the $nn$ pair from this level
will result in a $^{10}$C nucleus in the ground state. Taking into
account the statistical uncertainty in the number of experimental
events in the 2.0--3.0 MeV energy window (276$\pm$17), we
calculate $S_{\lim}=297$ at 90\% C.L. Accounting for the
associated efficiency $\varepsilon _{\Delta E}=0.44$,
 the limiting value for
$^{10}$C decays is $D_{\lim}=6.8\cdot10^2$. With the values
of $N_{nucl}=1.9\cdot 10^{29}$ for $^{12}$C nuclei and $t=29.1$ d,
we obtain
\begin{center}
$\tau_{\lim}(nn, ^{12}$C$)=4.4\cdot 10^{25}$ y with 90\% C.L.
\end{center}

For the $^{16}$O nucleus we will account for only one $nn$ pair on
the outermost $1p_{1/2}$ orbit ($nn$ decay in deeper levels will
result the $^{16}$O nucleus being too excited). With the number of
$^{16}$O nuclei (in a 1~m thick water layer) $N_{nucl}=9.8\cdot
10^{29}$, $N_{obj}=1$ and $D_{\lim}=1.4\cdot10^4$ for the energy
region 2.2--2.6 MeV, the result is

\begin{center}
$\tau_{\lim}(nn, ^{16}$O$)=5.7\cdot 10^{24}$ y with 90\% C.L.
\end{center}

With the assumption that the mean lifetime of the $nn$ pair is the same in $^{12}$C
and $^{16}$O nuclei, and that $nn$ decays in both of them contribute to the
experimental spectrum simultaneously, one can obtain a slightly more stringent
limit for $nn$ decay:
\begin{center}
$\tau_{\lim}(nn \rightarrow inv)=4.9\cdot 10^{25}$ y with 90\% C.L.
\end{center}

For $n$ decay in $^{12}$C we used a similar approach,
just demanding that the simulated response function for $^{11}$C decay
should be equal to the experimental spectrum in the energy region
of 1.0--1.1 MeV (fig. 2). In this way the value $D_{lim}=3.4\cdot10^3$
was determined for the full number of $^{11}$C decay events. Together
with $N_{obj}=4$ (four neutrons on the $1p_{3/2}$ level), it gives
the following limit:
\begin{center}
$\tau_{\lim}(n, ^{12}$C$)=1.8\cdot 10^{25}$ y with 90\% C.L.
\end{center}

\subsubsection{$p$ and $pp$ disappearance}

As for the $p$ and $pp$ decays into invisible channels, the $p$
disappearance in $^{13}$C will result in $^{12}$B nuclei, the
$\beta^-$ decaying with high energy release $Q$=13.370 MeV. The
$pp$ decays in $^{13}$C will produce $^{11}$Be nuclei, also the
$\beta^-$ decaying with $Q$=11.508 MeV (with probability of decay
to the ground state of 57\%).

To estimate the $\tau _{\lim}$ for $p$ and $pp$ instabilities, we
use the fact that no candidate scintillation events were observed
in the CTF spectrum with energies higher than 4.5 MeV (fig. 2).

High energy release in the liquid scintillator can activate the
muon veto of the CTF, resulting in rejection of the event. We take
into account such a suppression of high energy tails in $\beta$
decays of $^{12}$B and $^{11}$Be using the probability $\eta(E)$
for identification of an event with energy $E$ in the LS by the
muon veto (fig. 3). The beta spectra of $^{12}$B and $^{11}$Be
without and with suppression by the muon veto are also shown in
fig. 3. The part of the $^{12}$B beta spectrum with $E\geq4.5$ MeV
reduced by a factor of $1-\eta(E)$ gives an integrated efficiency
$\varepsilon _{\Delta E} = 0.39$. With zero observed events (and
with the assumption of zero expected background), the limiting
value for the number of events is $S_{\lim}=2.44$ with 90\% C.L.
in accordance with the Feldman-Cousins procedure \cite{Fel98}
recommended by the Particle Data Group \cite{PDG00}. Thus we
arrive at the value $D_{\lim}=6.2$ for $^{12}$B decay. Together
with $N_{nucl}=2.1\cdot 10^{27}$ for parent $^{13}$C nuclei and
the number of $N_{obj}=4$ for protons in $1p_{3/2}$ orbit, we
obtain
\begin{center}
$\tau_{\lim}(p, ^{13}$C$)=1.1\cdot 10^{26}$ y with 90\% C.L.
\end{center}

\nopagebreak
\begin{figure}[htbp]
\begin{center}
\mbox{\epsfig{figure=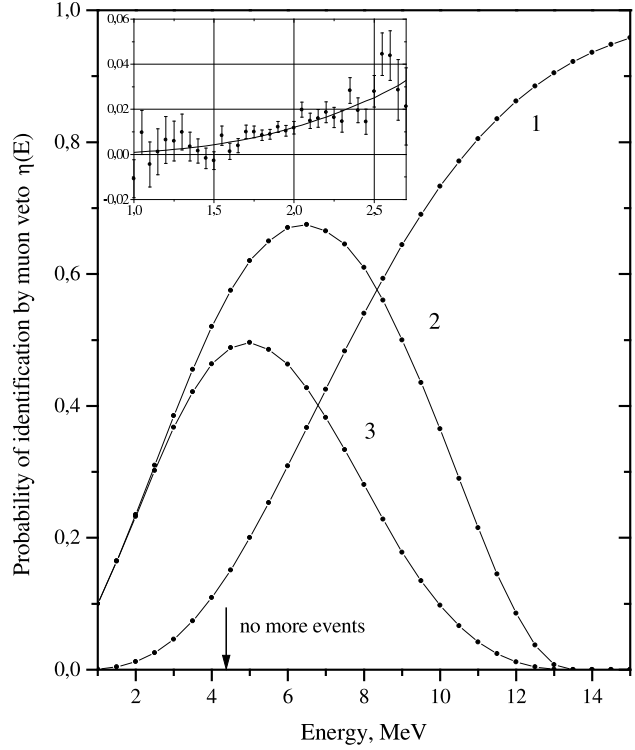,width=9.0cm}}
\caption{Probability of identification of an event with
energy $E$ in the scintillator by the muon veto (1). The $\beta$
spectra of $^{12}$B without (2) and with (3) suppression by the
muon veto are also shown in arbitrary units.
For signals with $E <$ $\simeq$3 MeV ($^{10}$C, $^{11}$C and
$^{14}$O decays), $\mu$ veto does not have a big effect on the overall
efficiency.
In the inset the experimental data taken with the radon source are presented.
The tagged events of $^{214}$Bi--$^{214}$Po at the energy $E$=1.9
MeV are 'seen' by the muon veto system with an efficiency of
$\eta=0.01$.}
\end{center}
\end{figure}

In a similar way, for the $pp$ decay in $^{13}$C with the value
$N_{obj}=2$ for the number of $pp$ pairs and
$\varepsilon_{\Delta E}=0.36$, the mean lifetime limit is
\begin{center}
$\tau_{\lim}(pp, ^{13}$C$)=5.0\cdot 10^{25}$ y with 90\% C.L.
\end{center}

All mean lifetime limits obtained here together with the numbers of parent nuclei,
$N_{obj}$ and the numbers of decay events are summarized in table 3.

\begin{table}[htbp]
\begin{center}
\caption{Mean lifetime limits, $\tau_{\lim}$, (at 90\% C.L.) for $N$
and $NN$ decays in the CTF. $N_{nucl}$ is the number of parent
nuclei; $N_{obj}$ the number of objects ($n$, $p$ and $NN$ pairs)
per parent nucleus; $D_{\lim}$ the excluded number of decay
events.}
\begin{tabular}{|lllclc|}
\multicolumn{6}{c}{~} \\ \hline \multicolumn{2}{|c}{Decay} &
$N_{nucl}$ & $N_{obj}$ & $D_{\lim}$ & $\tau_{\lim}$, y  \\ \hline
~ & & & & & \\ $p$  & $^{13}_{~6}$C$\rightarrow$$^{12}_{~5}$B  &
2.1$\cdot$10$^{27}$        & 4 & 6.2  & $1.1\cdot10^{26}$ \\ $n$ &
$^{12}_{~6}$C$\rightarrow$$^{11}_{~6}$C  & 1.9$\cdot$10$^{29}$ & 4
& $3.4\cdot10^3$  & $1.8\cdot10^{25}$ \\ $nn$ &
$^{12}_{~6}$C$\rightarrow$$^{10}_{~6}$C  & 1.9$\cdot$10$^{29}$ & 2
& $6.8\cdot10^2$   & $4.4\cdot10^{25}$ \\ ~    &
$^{16}_{~8}$O$\rightarrow$$^{14}_{~8}$O  &
9.8$\cdot$10$^{29}$$^{~a}$ & 1 & $1.4\cdot10^4$  &
$5.7\cdot10^{24}$ \\ $pp$ &
$^{13}_{~6}$C$\rightarrow$$^{11}_{~4}$Be & 2.1$\cdot$10$^{27}$ & 2
& 6.7  & $5.0\cdot10^{25}$ \\ ~ & & & & &
\\ \hline
\multicolumn{6}{l}{$^a$ In 1 m thick layer of water closest to the CTF liquid scintillator} \\
\end{tabular}
\end{center}
\end{table}

\section{Conclusions}

Using the unique features of the BOREXINO Counting Test Facility
-- the extremely low background, the large scintillator mass of 4.2 ton and
the low energy
 threshold -- new limits on $N$ and $NN$ decays into
invisible channels (disappearance, decays to neutrinos, majorons,
etc.) have been set:

\noindent
$\tau (n\rightarrow inv)>1.8\cdot
10^{25}$ y,
\noindent
$\tau (p\rightarrow inv)>1.1\cdot 10^{26}$ y,

\noindent
$\tau (nn\rightarrow inv)>4.9\cdot 10^{25}$ y
\noindent
and
\noindent
$\tau
 (pp\rightarrow inv)>5.0\cdot 10^{25}$ y with 90\% C.L.

Comparing these values with the data of table 1, one can see that
the established limits for $nn$ and $pp$ decays are the best
up-to-date limits set by any method, including radiochemical and
geochemical experiments.

  These limits are obtained in a very conservative assumption that
all events in the corrresponding energy region are due to the nucleon
decays. The data from the full scale Borexino detector will improve
the presented limits by at least two orders of magnitude.


\end{document}